\documentclass[12pt]{iopart}
\usepackage{iopams}  
\usepackage{graphicx}

\begin{document}
\title{Excited-state spectroscopy of single NV defect in diamond using optically detected magnetic resonance}
\author{P~Neumann$^{1}$, R~Kolesov$^{1}$, V~Jacques$^{1}$$^{\ast}$, J~Beck$^{1}$, J~Tisler$^{1}$, A~Batalov$^{1}$, L Rogers$^{2}$, N B Manson$^{2}$, G~Balasubramanian$^{1}$, F~Jelezko$^{1}$ and J~Wrachtrup$^{1}$}

\vspace{1cm}

\address{$^{1}$
Physikalisches Institut, Universit$\ddot{\rm a}$t Stuttgart, Pfaffenwaldring 57, D-70550 Stuttgart, Germany}
\address{$^{2}$ Laser Physics Centre, Australian National University, Canberra,
ACT 0200, Australia}

\address{$^{\ast}$ Corresponding author v.jacques@physik.uni-stuttgart.de}
\begin{abstract}
Using pulsed optically detected magnetic resonance techniques, we directly probe electron-spin resonance transitions in the excited-state of single nitrogen-vacancy (NV) color centers in diamond. Unambiguous assignment of excited state fine structure is made, based on changes of NV defect photoluminescence lifetime. This study provides significant insight into the structure of the emitting $^{3}\rm E$ excited state, which is invaluable for the development of diamond-based quantum information processing.
\end{abstract}


\submitto{\NJP}

\maketitle

Over the last decade, the negatively charged nitrogen-vacancy (NV) color center in diamond has attracted a lot of interest because it can be optically addressed as single quantum system~\cite{Gruber_Science1997} and exhibits several important properties for quantum information science applications. First, its perfect photostability at room temperature enables to realize a practical NV-based single photon source~\cite{Kurtsiefer_PRL2000,Beveratos_OptLett2000} for quantum cryptography applications~\cite{Beveratos_PRL2002,Alleaume_NJP2004}. Second, NV color centers have a paramagnetic ground state which spin can be optically polarized, read-out and exhibits long coherence time even at room temperature~\cite{Jelezko_PRL2004,Gaebel_NatPhys2006}. Coherent manipulation of electron and nuclear spins of single NV color centers has been used to realize solid-state quantum physics experiments, ranging from coherent coupling of a single NV color center to other single spins in the diamond crystalline matrix~\cite{Gaebel_NatPhys2006,Hanson_PRL2006,Lukin_Science2006}, to the implementation of a quantum register~\cite{Lukin_Science2007} and conditional two-qubit CNOT gates~\cite{Jelezko_PRL2004bis}, and very recently the generation of Bell and GHZ states with long coherence times~\cite{Neumann_Science2008}.\\
\indent Despite these results, which make the NV color center a competitive candidate for solid-state quantum information processing, the excited-state structure of the defect is not yet fully understood~\cite{Manson_PRB2006,Manson_JLum2007}. This knowledge is however crucial for single-spin high-speed coherent optical manipulation through $\Lambda$-based transitions~\cite{Hemmer_OptLett2001,Santori_OptExp2006,Santori_PRL2006} as well as for future implementation of quantum information protocols like quantum repeaters~\cite{Childress_PRL2006,Childress_PRA2006} that can be used as a building block for a quantum network~\cite{Cirac_PRL1997}.\\
\indent Resonant optical excitation of single NV color centers at low temperature~\cite{Tamarat_NJP2008} and cw electron-spin resonance experiments~\cite{Afshalom_QuantPh2008} have recently provided new insights into the structure of the excited-state, showing that its fine structure is strongly affected by local strain in the diamond matrix~\cite{Tamarat_NJP2008,Afshalom_QuantPh2008}. Recent ensemble experiments have also studied the behaviour of an infrared emission line that gives a better understanding of the metastable state responsible for spin polarization~\cite{Rogers_CondMat2008}.\\
\indent Here we develop a new approach to probe the excited-state fine structure of single NV color centers. Using pulsed optically detected magnetic resonance techniques, we directly probe electron-spin resonance transitions of the excited-state. Unambiguous assignment of excited state fine structure is made, based on changes of NV defect photoluminescence lifetime. 
\\

\indent The NV defect center in diamond consists of a substitutional nitrogen atom (N) associated with a vacancy (V) in an adjacent lattice site of the diamond crystalline matrix (Fig. 1-(A)). For the negatively charged NV color center, which is addressed in this study, the ground state is a spin triplet state $^{3} \rm A$, originating from six unpaired electron spins. Owing to $\rm C_{3v}$ symmetry of the NV center, ground state spin sublevels $m_{s}=\pm 1$ are degenerate and the zero-field splitting from $m_{s}=0$ is $D_{gs}=2.87$ GHz (Fig. 1-(B)). The excited state $^{3} \rm E$ is also a spin triplet, associated with a broadband photoluminescence emission with zero phonon line at $1.945$ eV. The order of other energy levels is still under debate but it is now well established that at least one metastable state $^{1}\rm A$ is lying between the ground and excited triplet state~\cite{Manson_PRB2006,Rogers_CondMat2008}. This metastable state plays a crucial role in spin dynamics of the NV color center. Indeed, whereas the optical transitions $^{3} \rm A\rightarrow ^{3}$E are spin conserving, non-radiative inter-system crossing transitions to the metastable state $^{3} \rm E\rightarrow^{1}$A are strongly spin selective as the shelving rate from the $m_{s}=0$ sublevel is much smaller than those from $m_{s}=\pm 1$. Conversely, the metastable state decays preferentially towards the ground state $m_{s}=0$ sublevel, leading to a strong spin polarization into $m_{s}=0$ after a few optical excitation-emission cycles~\cite{Jelezko_APL2002}.
\begin{figure}[t]
\centerline{\resizebox{0.7\columnwidth}{!}{
\includegraphics{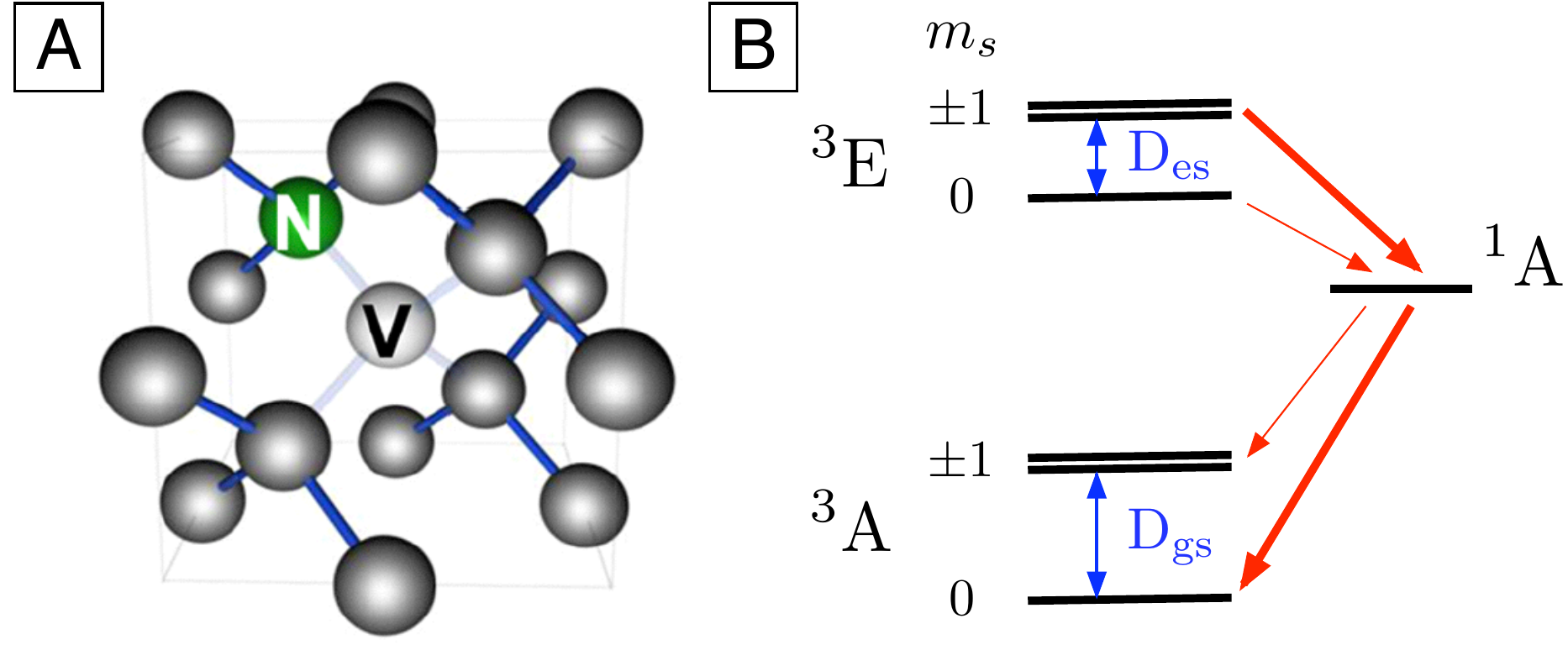}}}
\caption{A-Atomic structure of the nitrogen-vacancy (NV) centre. The single substitutional nitrogen atom (N) is accompanied by a vacancy (V) at a nearest neighbour lattice position. B-Simplified energy-level diagram of the NV center. $D_{gs}$ and $D_{es}$ correspond respectively to the zero-field splitting between $m_{s}=0$ and $m_{s}=\pm1$ in the triplet ground state $^{3} \rm A$ and in the triplet excited state $^{3} \rm E$. Spin selective shelving into a metastable singlet state $^{1}\rm A$ (red arrows) leads to spin-polarization of the centre into the ground state $m_{s}=0$ sublevel through optical pumping.}
\end{figure}

\indent  Another consequence of this spin-selective process is that the photoluminescence intensity is higher when the $m_{s}=0$ state is populated, allowing optical detection of spin-rotation of a single NV center at room temperature by optically detected magnetic resonance (ODMR)~\cite{Wrachtrup_ODMR,Moerner_ODMR}. Indeed, if a single NV center, initially prepared in the $m_{s}=0$ state through optical pumping, is driven to the $m_{s}=\pm 1$ spin sublevels by applying a resonant microwave frequency, a decrease in photoluminescence signal is observed. This technique is now routinely used for single-spin readout in solid-state quantum optics experiments using single spins in diamond as quantum bits. Until now only the ground state spin sublevels have been detected using ODMR. However, as optical transitions are spin conserving, spin rotations in the excited states should be also detected, allowing to probe the structure of spin sublevels.\\

\indent We first investigate single NV color centers artificially created in a ultra-pure type IIa diamond sample (Element6), by implanting $7$ MeV isotopically pure $^{15}\rm N$ atoms and by annealing the sample for two hours in vacuum at $800 \ ^{\circ}\rm C$~\cite{Rabeau_APL2006}. NV centers are then optically addressed at room temperature using a standard confocal microscope coupled to a Hanbury-Brown and Twiss setup used to measure the photoluminescence second order corrrelation function $g^{(2)}(\tau)$ and verify that an individual NV center is adressed.
\begin{figure}[h!]
\centerline{\resizebox{1.0\columnwidth}{!}{
\includegraphics{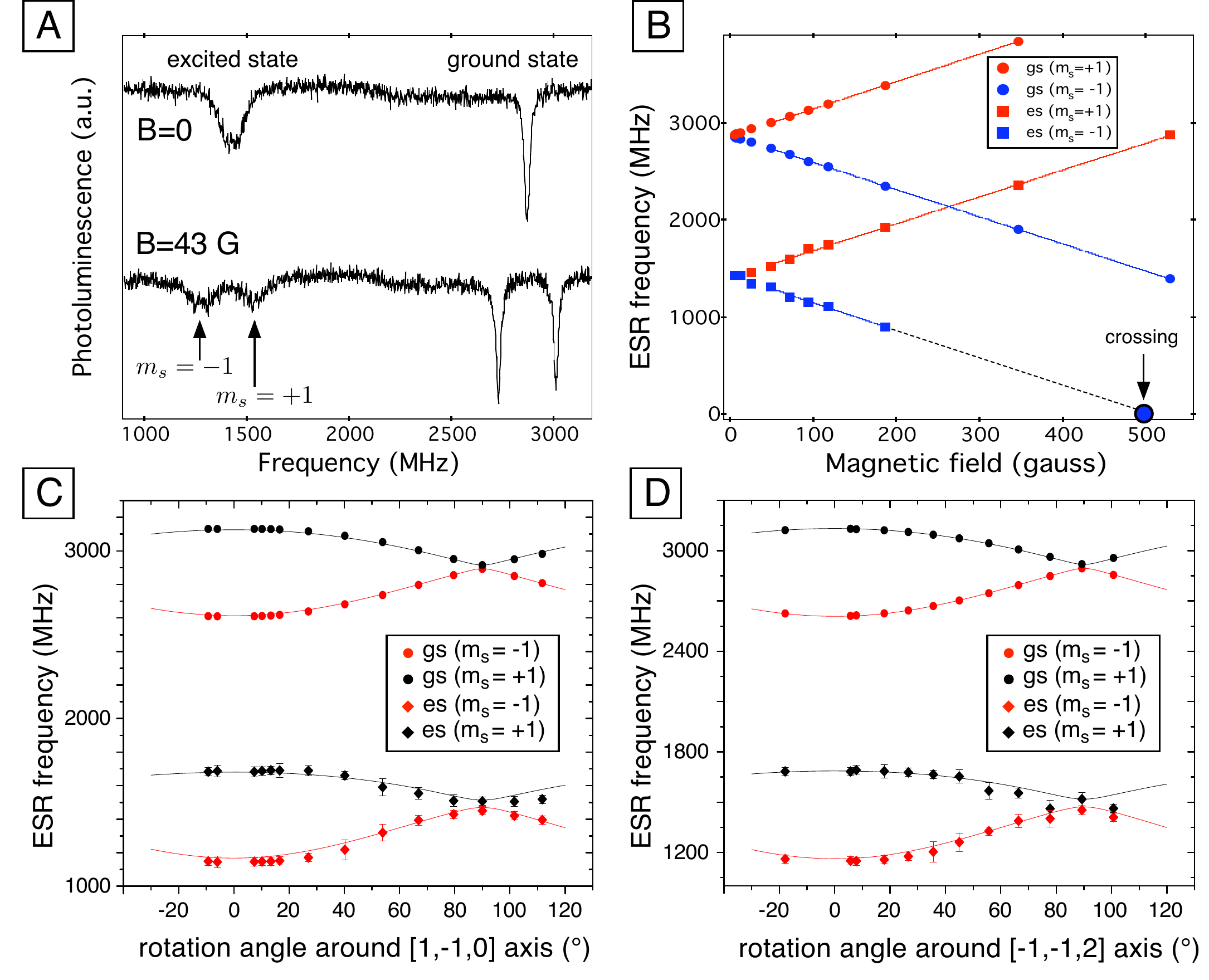}}}
\caption{A-ODMR spectra of a single NV color center at zero magnetic field (upper trace) and with a magnetic field of amplitude $B=43$ G applied along the NV symmetry axis which corresponds to a [111] crystal axis (bottom trace). Electron spin resonances are evidenced both in the ground state (gs) and in the excited state (es). B-Electron spin resonance frequencies as a function of the amplitude of the applied magnetic field along a [111] crystal axis. Solid lines correspond to a fit of the experimental results using equation~(\ref{ESR}), leading to $D_{es}=1423\pm10$ MHz and $g_{es}=2.01\pm 0.08$. A level anti-crossing between the excited state $m_{s}=0$ and $m_{s}=-1$ sublevels is expected for a magnetic field aroung $500$ G. C and D-Electron spin resonance frequencies measured by rotating a magnetic field of amplitude $B=92$ G around a [1,-1,0] crystal axis (C) and around a [-1,-1,2] crystal axis (D). The solid lines correspond to theoretical expectations using the Hamiltonian described by equation~(\ref{Hamilto}) with the measured values of $D_{es}$ and $g_{es}$, and neglecting the strain-induced splitting $E_{es}$.}
\end{figure}

\indent Electron spin resonance (ESR) spectroscopy of single NV centers is realized by applying microwaves, using a copper microwire ($20 \ \mu$m diameter) close to the NV center, and by monitoring the photoluminescence intensity. When the microwave frequency is resonant with the transition between $m_{s}=0$ and $m_{s}=\pm1$ sublevels, spin rotation is evidenced as a dip of the photoluminescence signal as explained above. Figure 2-(A) (upper trace) describes a typical ODMR signal obtained by sweeping the microwave frequency without any applied magnetic field. The well-known ground state transition between $m_{s}=0$ and $m_{s}=\pm1$ sublevels is detected at $2.87$ GHz, and an additional broad line around $1.4$ GHz is observed. By applying a magnetic field to the sample, the degeneracy of $m_{s}=\pm 1$ is lifted by the Zeeman effect, leading to the appearance of two lines in corresponding resonances of the ODMR spectrum (see figure 2-(A) bottom trace). As an hypothesis, we attribute the broad resonance to spin sublevels of an excited state of the NV color center. In the following, we will demonstrate that this excited state actually corresponds to the emitting excited state $^{3}\rm E$.\\
\indent We first study in more details the ESR frequency positions as a function of the magnitude of a magnetic field ($B$) applied along the NV symmetry axis which corresponds to a [111] crystal axis. The results of this experiment are depicted on figure 2-(B). Neglecting electron-nuclear spin coupling, the excited-state spin Hamiltonian of the NV defect can be written as:
\begin{equation}
\label{Hamilto}
H=D_{es}(S^{2}_{z}-\frac{2}{3})+E_{es}(S^{2}_{x}-S^{2}_{y})+g_{es}\mu\vec{B}\cdot\vec{S} \ ,
\end{equation}
where $D_{es}$ is the excited-state zero-field splitting, $S=1$, $E_{es}$ is the excited-state strain-induced splitting coefficient, $g_{es}$ the excited state g-factor and $\mu$ the Bohr magneton.  By considering magnetic field magnitudes such that the strain-induced fine structure splitting is negligible compared to Zeeman splitting ($\left|E_{es}(S^{2}_{x}-S^{2}_{y})\right| \ll \left|g_{es}\mu\vec{B}\cdot\vec{S}\right|$), the resonant frequencies $\omega_{\pm}$ associated to eigenstates $m_{s}=\pm1$ are given by 
\begin{equation}
\label{ESR}
\omega_{\pm}=D_{es}\pm g_{es}\mu B \ .
\end{equation}

\indent For the same NV center considered in this study, it was not possible to observe any strain-induced splitting in the excited state, the ODMR dip being very broad with a FWHM on the order of $100$ MHz (see figure 2-(A)). As a result, it is reasonable to consider that the strain-induced splitting coefficient is such that $2E_{es} \ll 100$ MHz~\cite{Strain}. Following this consideration the experimental results depicted on figure 2-(B) can be fitted using equation~(\ref{ESR}) for magnetic field magnitudes bigger than $50$ G, corresponding to a Zeeman splitting on the order of $130$ MHz ($\gg E_{es}$). The results of the fit lead to $D_{es}=1423\pm10$ MHz and an isotropic g-factor $g_{es}=2.01\pm 0.08$ which is similar to the ground state g-factor. This isotropy indicates that the orbital angular momentum does not play a significant role in the excited state. Finally, the positions of ESR frequencies were measured by rotating a magnetic field of magnitude $B=92$ G around a [1,-1,0] crystal axis (figure 2-(C)) and around a [-1,-1,2] crystal axis (figure 2-(D)). The experimental results provide a strong evidence that the ground and excited states exhibit the same orientations. \\
\indent It is interesting to note that a level anti-crossing between the $m_{s}=0$ and $m_{s}=-1$ sublevels of the excited state is expected for a magnetic field amplitude on the order of $500$ G (see figure 2-(B)). This explains why a decrease of the NV color center photoluminescence has been observed in ensemble experiments at such magnetic field~\cite{Rogers_CondMat2008,Epstein_NatPhys2005}. Indeed, when a level anti-crossing occurs in the excited state, the electron spin polarization of the center is significantly reduced. This effect is identical to the well-known photoluminescence dip occuring at $B=1028$ G, when the $m_{s}=0$ and $m_{s}=-1$ sublevels of the ground state are crossing.\\
\indent The cw experiments reported above prove that the ESR signal corresponds to an excited state, but do not provide ultimate proof that this state is responsible for fluorescence emission. For example, other ÒdarkÓ state involved in spin polarization pathway can influence the spin polarization of NV defect~\cite{Rogers_CondMat2008}. We now demonstrate that the excited state observed in ODMR spectra actually corresponds to the emitting excited state $^{3}\rm E$. Following a method introduced in Ref.~\cite{Batalov_PRL2008}, the experiment is based on observing a modification of photoluminescence decay by manipulating the excited state spin sublevels with resonant microwave pulses.
\begin{figure}[b]
\centerline{\resizebox{0.70\columnwidth}{!}{
\includegraphics{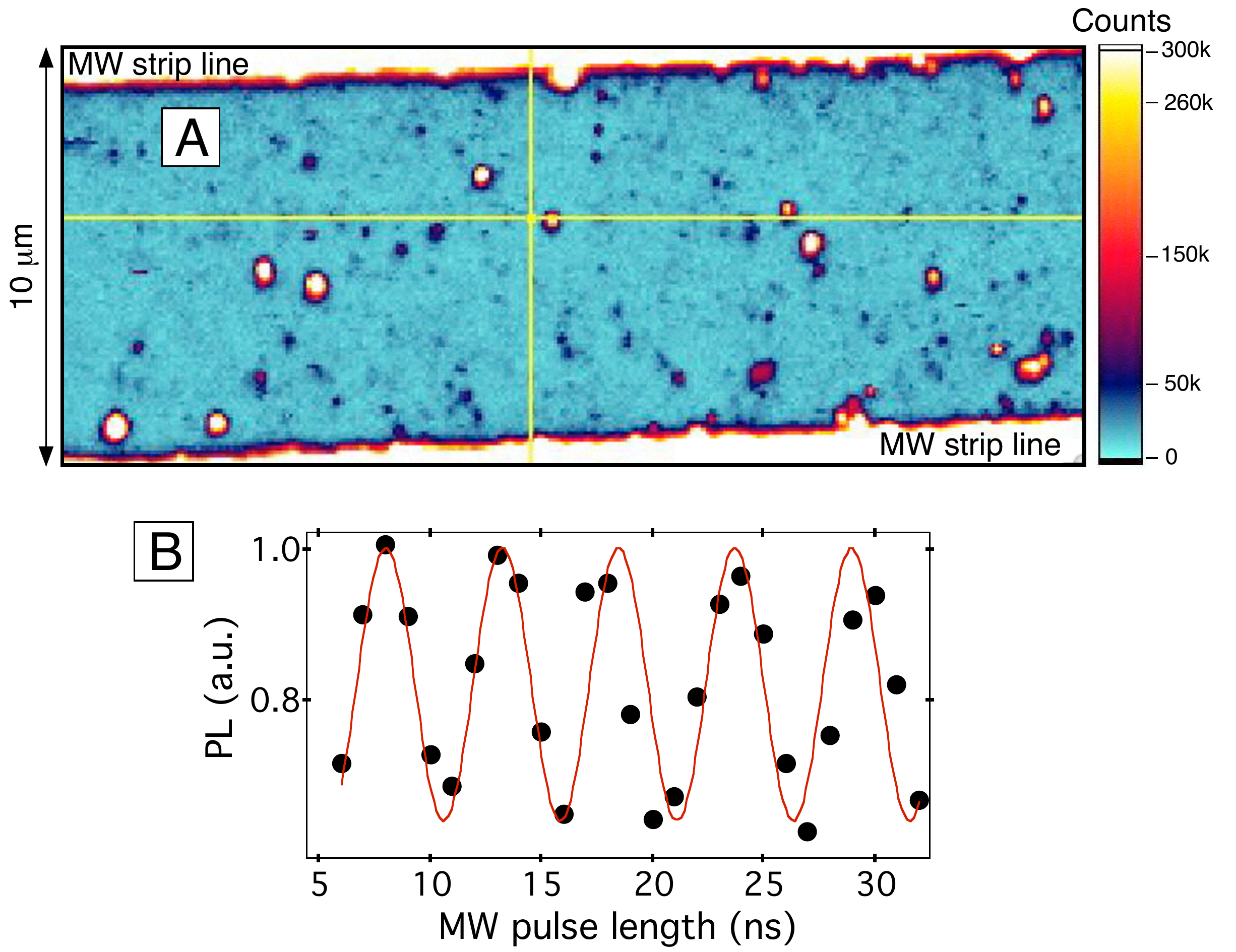}}}
\caption{A-Confocal fluorescence image of the sample showing strip-line microwave wires and NV centers in diamond nanocrystals. Fluorescence intensity is encoded in 
colour scale and the yellow cursors indicate the NV center of interest, which was verified as a single emitter using photon correlation measurements (data not shown). B-Rabi nutation between ground state spin levels of this single NV center, indicating a $2\pi$ rotation within $5$ ns.}
\end{figure}

\indent For that purpose, a large microwave driving field is required in order to reach a significant change in the population of the excited-state spin sublevels within the radiative lifetime. In order to more easily meet that requirement, we used single NV color centers in diamond nanocrystal, for which the photoluminescence decay is known to be much longer than in bulk diamond sample~\cite{Beveratos_PRA2001}. In addition, the nanodiamonds were spin-coated on a microscope cover glass on which gold strip-line microwave wires had been deposited using shadow mask photolithography and metal electrodepositing in order to reach high ESR Rabi frequencies. Typical dimensions of the wires were $10 \ \mu$m width and $2 \ \mu$m thick. Figure 3-(A) shows a photoluminescence map of the sample, the yellow cursors indicating the individual NV center studied in the following. For this NV center, we measured a Rabi nutation between ground state spin levels at a frequency of $200$ MHz, corresponding to a $\pi$ pulse of $2.5$ ns (figure 3-(B)).\\
\indent As depicted on figure 4-(A), the ODMR spectrum shows a huge energy splitting between ESR lines, even at small magnetic field magnitude ($B=20$ G). Such splitting, which can not be explained by the Zeeman effect, results from strain-induced splitting which is known to be much stronger in diamond nanocrystals than in bulk samples~\cite{Jelezko_JMol2001}. However, note that strain has almost no effect on the $D$ values ($D_{gs}$ and $D_{es}$) whereas it does cause the $E$ values ($E_{gs}$ and $E_{es}$ in the Hamiltonian described by equation~(\ref{Hamilto})) to become non-zero. In the following, we use resonant microwaves with the ground ($\rm MW_{gs}$) and excited state ($\rm MW_{es}$) $m_{s}=0\rightarrow m_{s}=-1$ transitions, at $2844$ MHz and $1000$ MHz respectively (see figure 4-(A)).\\
\indent As a first step, the excited state lifetime associated with each spin sublevels was measured. The NV center was first polarized into the ground state $m_{s}=0$ sublevel using an optical pulse of duration $3 \ \mu$s at the wavelength $\lambda=532$ nm. After a time delay of $1 \ \mu$s, which ensures that the NV center has relaxed to the ground state, an optical pulse (40 ps, $\lambda=532$ nm) much shorter than the radiative lifetime was used to excite the NV center. As the optical transition is spin-conserving, such a sequence allows to build up the photoluminescence decay of the excited-state $m_{s}=0$ sublevel using a standard start-stop technique for lifetime measurements, the start being a part of the pulsed excitation and the stop being the single-photon counter signal. For measuring the decay of $m_{s}=-1$ sublevel, an additional microwave $\pi$ pulse resonant with ground state spin transition ($\rm MW_{gs}$) was introduced before the ps optical pulse.
\begin{figure}[t]
\centerline{\resizebox{1.0\columnwidth}{!}{
\includegraphics{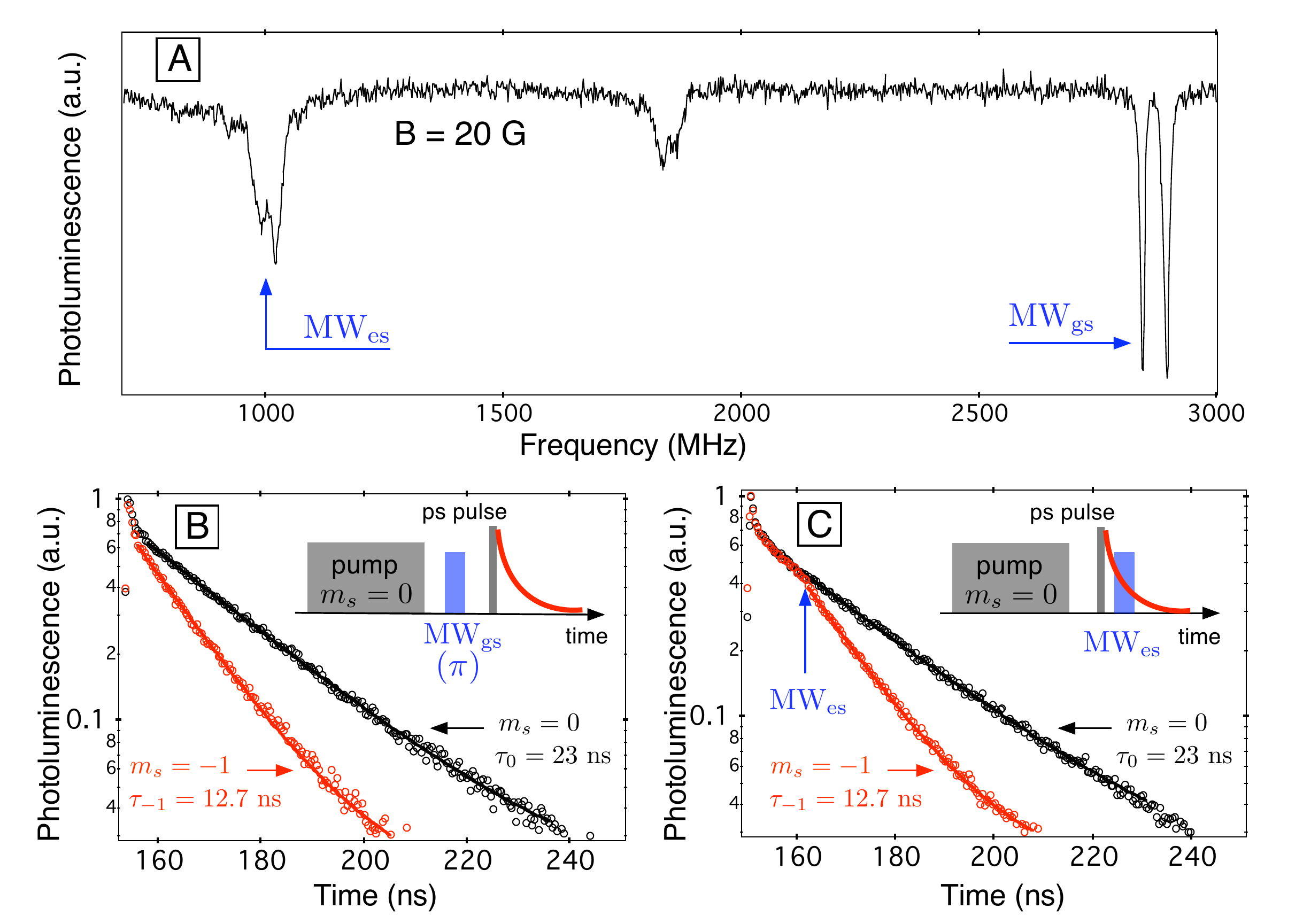}}}
\caption{A-ODMR spectra of the studied single NV color center in a diamond nanocrystal recorded with a $20$ G magnetic field applied. $\rm MW_{gs}$ and $\rm MW_{es}$ respectively correspond to the resonant frequency of $m_{s}=0\rightarrow m_{s}=-1$ transition in the ground state and in the excited state. Note that careful examination of the excited state resonances shows multiple dips which might be related to excited-state hyperfine structure~\cite{Afshalom_QuantPh2008}. B and C-Spin-selective photoluminescence decay curves, monitored by applying spin selective resonant microwave pulse (B) in the ground state (frequency $\rm MW_{gs}$) and (C) on the excited state (frequency $\rm MW_{es}$). The insets describe the time sequence of the experiments (see main text) and solid lines are single exponential fit curves, leading to a  radiative lifetime $\tau_{0}=23$ ns  for $m_{s}=0$ sublevel and  $\tau_{-1}=12.7$ ns for $m_{s}=-1$.}
\end{figure}

\indent The results of such measurements are depicted on figure 4-(B). Fluorescence decays follow single exponential decay associated to a lifetime $\tau_{0}=23$ ns  for $m_{s}=0$ sublevel and  $\tau_{-1}=12.7$ ns for $m_{s}=-1$. As expected, these values are much bigger than the ones measured in bulk samples~\cite{Batalov_PRL2008}. As the photoluminescence is always smaller from $m_{s}=\pm1$ sublevels compared to $m_{s}=0$, the measured lifetime difference results from spin-selective non-radiative inter-system crossing transition to the metastable state, and not from a modification of the transition strength. It is interesting to notice that the measured lifetime ratio, $\tau_{-1}/\tau_{0}\approx 0.55$, is in good agreement with recent theoretical predictions~\cite{Manson_PRB2006} and previously reported measurements in bulk samples~\cite{Batalov_PRL2008}.\\
\indent In order to check if the resonance lines observed in ODMR spectra actually correspond to spin transitions in the emitting $^{3}\rm E$ excited state, the same experiment is performed by applying a microwave pulse resonant with the excited state spin transition $m_{s}=0\rightarrow m_{s}=-1$ ($\rm MW_{es}$), just after the picosecond optical pulse. The results, depicted in figure 4-(C), indicate a drastic change of the photoluminescence decay when the microwave pulse is applied. Before the microwave excitation, the photoluminescence decay follows the single exponential decay associated with the excited-state $m_{s}=0$ sublevel ($\tau_{0}=23$ ns). The microwave excitation then suddenly rotates the spin in the excited state leading to the exponential decay associated with the excited-state $m_{s}=-1$ ($\tau_{-1}=12.7$ ns). These results unambigously evidence that the new ESR lines observed in ODMR spectra are related to fine structure of the emitting excited state $^{3}\rm E$.\\
\indent It was not possible to detect Rabi nutations on the excited state by varying the resonant microwave pulse duration. This appears as a difficult issue because of many factors. Among them are the large width of the excited-state ESR resonance, a strong hyperfine coupling owing to high spin density of the excited-state wavefunction at the nitrogen nucleus~\cite{Gali_PRB2008} and the short radiative lifetime of NV color centers.\\
\indent Finally, we would like to shortly discuss our observations in the context of previously reported models of the excited-state structure. Resonant optical excitation of single NV color centers at low temperature has recently indicated that the excited-state is actually an orbital doublet, split into two orbital singlet branches by local strain~\cite{Tamarat_NJP2008}. Then, we would expect to observe four excited-state resonances when a magnetic field is applied to the NV center, while we only ever observed two lines (figures 2-(A) and 4-(A)). We tentatively identify the observed excited-state features to the upper branch because this branch shows high difference in inter-system crossing rate to the metastable state~\cite{Manson_PRB2006,Batalov_PRL2008}, as observed in experiments (see figure 4-(B) and (C)). However, previous model of the excited-state structure has predicted non vanishing $E_{es}$ in the upper branch~\cite{Tamarat_NJP2008}. The observations in bulk dimond reported in this paper ($E_{es}\approx 0$) bring then into question the completeness of currently available models for NV center excited-state structure.\\

\indent Summarizing, we have performed the excited-state spectroscopy of single NV color centers in diamond using cw and pulsed ESR techniques. This work provides significant insight into the structure of the emitting $^{3}\rm E$ excited state, which might be useful for diamond-based quantum information processing using $\Lambda$-based transitions for high-speed coherent optical manipulation of single spins~\cite{Santori_PRL2006} as well as for entanglement protocols used in quantum repeaters~\cite{Childress_PRL2006,Monroe_Nature2007}.

\ack{We acknowledge financial support by the European Union (QAP, EQUIND, and NEDQIT), Deutsche Forschungsgemeinschaft (SFB/TR21) and Australian Research Council. V J is supported by the Humboldt Foundation.}

\end{document}